\documentclass[runningheads]{llncs}
\usepackage[T1]{fontenc}
\usepackage{graphicx}
\usepackage{subcaption}
\usepackage{array}
\usepackage{booktabs}
\usepackage{listings}
\usepackage{hyperref}
\usepackage{subcaption} 
\usepackage[export]{adjustbox}

\let\subparagraph\relax
\usepackage{titlesec}
\titlespacing{\subsubsection}{0em}{0.5em}{0.5em}

\begin{document}

\newcommand{\github}{\url{https://github.com/zqh0421/slideitright}}
\title{
SlideItRight: Using AI to Find Relevant Slides and Provide Feedback for Open-Ended Questions
}

%
\titlerunning{SlideItRight}
%


\author{
Chloe Qianhui Zhao\inst{1} \and
Jie Cao\inst{2}  \and
Eason Chen\inst{1} \and
Kenneth R. Koedinger\inst{1} \and
Jionghao Lin\inst{3,1, }\thanks{Corresponding author.}
}
%
\authorrunning{C. Q. Zhao et al.}
%
\institute{
    Carnegie Mellon University, Pittsburgh PA 15213, USA \\
    \email{{\{cqzhao, easonc13, koedinger, jionghao\}@cmu.edu}} \and
    University of Pittsburgh, Pittsburgh PA 15260, USA \\
    \email{jic319@pitt.edu} \and
    The University of Hong Kong, Pokfulam Rd, Hong Kong, China\\
    \email{jionghao@hku.hk} 
}
\maketitle              
\begin{abstract} 
\vspace{-5mm}

Feedback is important in supporting student learning. While various automated feedback systems have been implemented to make the feedback scalable, many existing solutions only focus on generating text-based feedback. As is indicated in the multimedia learning principle, learning with more modalities could help utilize more separate channels, reduce the cognitive load and facilitate students' learning. Hence, it is important to explore the potential of Artificial Intelligence (AI) in feedback generation from and to different modalities. Our study leverages Large Language Models (LLMs) for textual feedback with the supplementary guidance from other modality - relevant lecture slide retrieved from the slides hub. Through an online crowdsourcing study (N=91), this study investigates learning gains and student perceptions using a 2×2 design (i.e., human feedback vs. AI feedback and with vs. without relevant slide), evaluating the clarity, engagement, perceived effectiveness, and reliability) of AI-facilitated multimodal feedback. We observed significant pre-to-post learning gains across all conditions. However, the differences in these gains were not statistically significant between conditions. The post-survey revealed that students found the slide feedback helpful in their learning process, though they reported difficulty in understanding it. Regarding the AI-generated open-ended feedback, students considered it personalized and relevant to their responses, but they expressed lower trust in the AI feedback compared to human-generated feedback.

\vspace{-2mm}

\end{abstract}
\vspace{-8mm}
\section{Introduction}
Providing feedback is widely acknowledged as crucial and effective for student learning~\cite{hattie_power_2007,Ryan18082021,wisniewski_power_2020,Yang01042013}. However, high student-teacher ratios often make it difficult for teachers to offer the level of individualized attention that diverse learners need. In particular, incorporating multiple modalities (such as text, visuals, and interactive elements) into feedback further complicates this task \cite{chen2023gptutor,hattie_power_2007}, which can cause learners to receive generic or insufficient guidance, reducing both engagement and learning outcomes.

\begin{figure}[t!]
    \centering
    \includegraphics[width=0.88\linewidth]{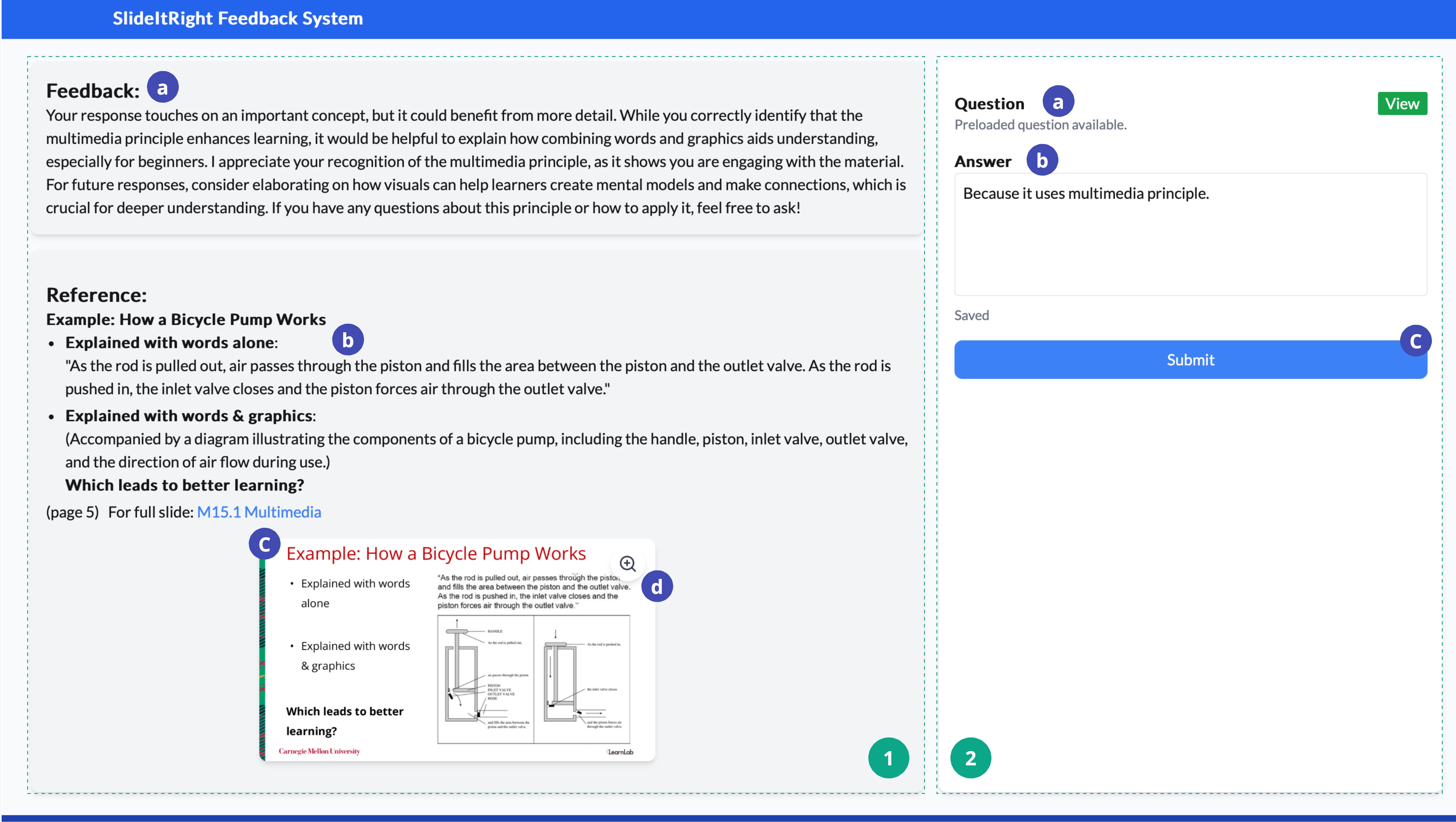}
    \caption{SlideItRight Student Interface and Functionality Overview. On the left is (1) the multimodal feedback panel, which includes (a) AI generated feedback on student responses, (b) OpenAI vision's understanding of the retrieved slide page, (c) a slide page related to the preset question, and (d) a zoom in button to enlarge and look through the retrieved slide page. On the right is (2) the user interaction panel, which features: (a) a preset question whose interface is currently minimized for Intelligent Tutoring System (ITS) platform integration, (b) student answer input box, (c) a status indicator of whether the student’s input is successfully cached.}
    \label{fig:mufin-interface}

\vspace{-5mm}
\end{figure}

Recent advances in Artificial Intelligence (AI), particularly Large Language Models (LLMs), present promising pathways to address these challenges \cite{DAI2024100299}. AI tools like tutoring chatbots and automated question-generation systems demonstrate the capacity to provide human-like responses at scale \cite{Lamsiyah24,Ma2024,Taneja24}, paving the way for more adaptable multimodal feedback aligned to individual learner needs \cite{DAI2024100299,olney_improving_2024}. Despite this potential, it remains unclear how effectively LLM-generated feedback enhances students’ learning and how learners perceive them.

Clarifying these questions is essential for guiding the effective adoption of AI in the classroom. While learning gain measurements can validate the pedagogical value of new technologies \cite{Hao02012022,irons2021enhancing}, student perceptions also offer critical insights - learners with more positive attitudes generally adopt deeper learning approaches \cite{book,dweck2006mindset}. Nonetheless, limited research has simultaneously examined both learning gains and student experience in LLM-facilitated multimodal feedback environments \cite{WANG2024124167}. To address this gap, we pose two core \textbf{R}esearch \textbf{Q}uestions:

\noindent \textbf{RQ1:} How does AI-facilitated multimodal feedback impact learning gains?

\noindent \textbf{RQ2:} How do learners perceive the effectiveness, clarity, and usability of LLM-facilitated multimodal feedback in supporting their learning process?

\noindent Our research makes two primary contributions to address these questions:
\begin{enumerate}
    \item We implement the SlideItRight system (Fig. \ref{fig:mufin-interface}), which is an automated feedback system that integrates AI-generated textual feedback with retrieved relevant lecture slides to enhance learning support.
    \item Using a 2×2 crowdsourcing study, we examine how different feedback modalities affect learning outcomes and student perceptions in an online setting. Through pre/post-tests and surveys, we assess learning gains and perceptions of clarity, engagement, effectiveness, and ease of use.
\end{enumerate}
\vspace{-4mm}

\section{Related Work}
\subsubsection{Feedback in Learning} 

Feedback is a process to support student learning \cite{Ryan18082021,Yang01042013}, with extensive research showing its effectiveness in improving academic performance \cite{hattie_power_2007,wisniewski_power_2020}. Recent learner-centered feedback framework emphasizes that the feedback content should (1) strengthen student-teacher relationships, (2) provide corrective information on performance, and (3) offer clear guidance for improvements \cite{Ryan18082021}. This comprehensive approach has been widely employed in recent studies \cite{ALDINO2024100332,Liang2024,jlin_learner_centred} for evaluating the effectiveness of feedback. However, the multifaceted nature of learner-centered feedback 
presents challenges for teachers attempting to manually craft it effectively at scale. 


\subsubsection{Multimodal Feedback} 
As stated in Mayer's multimedia learning principle, people learn more deeply from multiple modalities together than from a single modality, as learners process multimodal information through separate channels with limited capacity \cite{MAYER200285}. Studying discourse through a single modality can oversimplify and distort the actual nature of pedagogical practices \cite{ohalloran2008,parcalabescu_what_2021}. Instead, multimodal feedback integrates multiple representational modes or communication channels to convey information through various semiotic resources, including gestures, digital sources, and interactive components \cite{jewitt_routledge_2011,kress_multimodal_2001}. However, the effectiveness of multimodal feedback faces a significant challenge in managing cognitive load, as the interaction of verbal and visual stimuli in corrective discourse does not automatically enhance learners' attention to corrected forms \cite{todd2012psychophysical}. The critical aspect lies in how different ``atomic units of information'' work together for specific tasks with careful integration and balance \cite{ohalloran2008,parcalabescu_what_2021}.

\subsubsection{Large Language Models for Feedback Generation}
Recent advances in LLMs have opened new possibilities to generate quality textual feedback at scale, offering promising solutions to the challenges of traditional feedback mechanisms \cite{KASNECI2023102274,li2024bringing}. Researchers have leveraged LLMs to analyze student responses, generate contextual explanations, and offer comprehensive learner-centered feedback, where LLMs can provide detailed explanations while maintaining scalability and consistency \cite{DAI2024100299,qinjin_jia_llm-generated_2024,Lin2024IJAIED}. However, these models also face issues like accuracy and ``hallucination'', \cite{KASNECI2023102274}. Consequently, researchers have been exploring approaches to mitigate these limitations through techniques such as fine-tuning, reinforcement learning, and Retrieval-Augmented Generation (RAG) \cite{han2024improving,Lamsiyah24,lewis_retrieval-augmented_2021}.

Taken together, these research threads indicate both the promise and complexity of delivering timely, personalized, and pedagogically sound multimodal feedback that supports student active learning and feedback-seeking processes. To address these challenges, we introduce SlideItRight, a retrieval-augmented feedback system that combines LLM-based textual explanations with relevant instructional slides, as detailed in the following section.

\vspace{-2mm}

\section{System Implementation}
\begin{figure}[b!]
    \centering
    \includegraphics[width=0.9\linewidth]{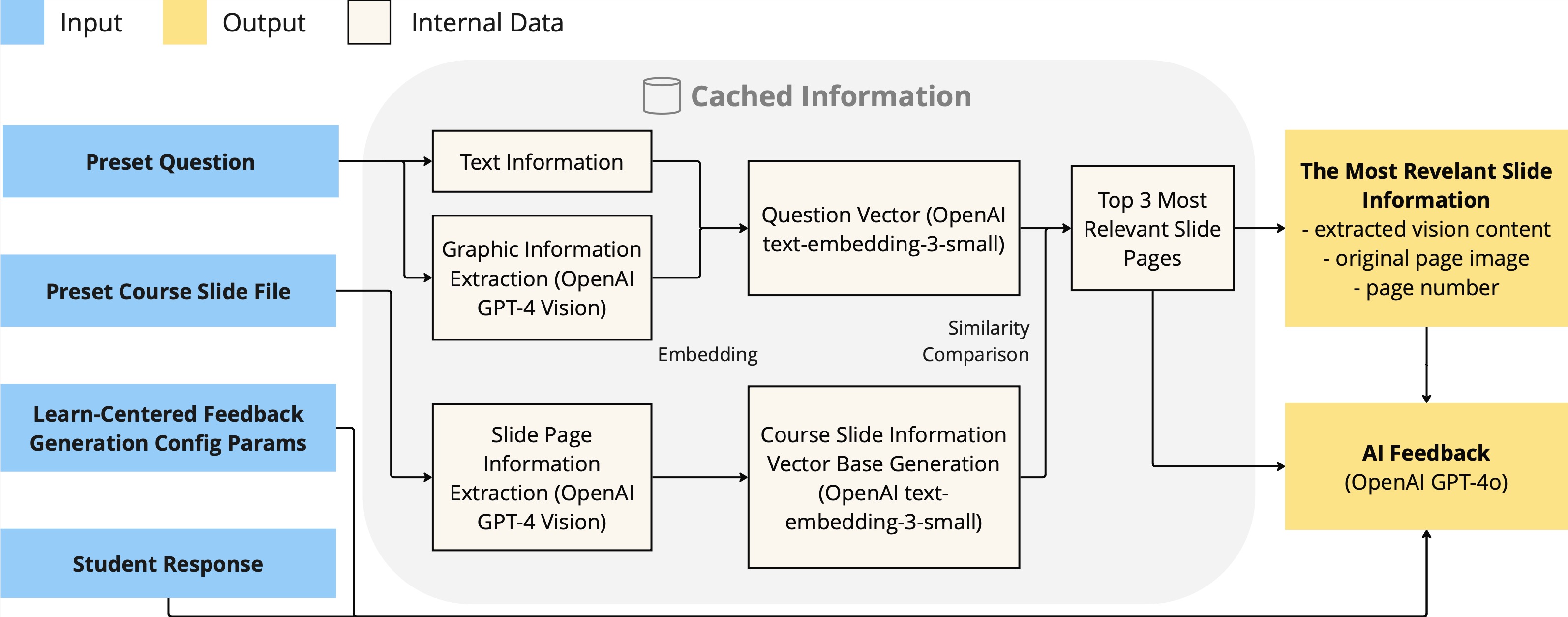}
    \caption{SlideItRight Workflow. The system retrieves relevant course slides based on the content of a given question, to enhance multimodal feedback generation for the student responses.}
    \label{fig:workflow}
\end{figure}

SlideItRight (Fig. \ref{fig:mufin-interface}) feedback system is designed to provide personalized multimodal feedback to support student learning through the integration of LLMs. The implementation focuses on providing effective feedback while creating an engaging learning experience for students. The workflow of our proposed feedback system is shown in Fig. \ref{fig:workflow}. The system's core functionality is built on four technological components: 1) multimodal input processing, 2) RAG-enhanced LLM feedback generation with retrieved lecture slide for reference, 3) feedback personalization, and 4) response time optimization, to address the primary challenges in delivering educational feedback. These technical capabilities work together to generate and deliver personalized feedback efficiently.


\subsubsection{Multimodal Input}
As shown in Fig. \ref{fig:workflow}, our system processes four types of multimodal inputs: system configuration parameters, student responses, question contents, and course slide files. Our system employs OpenAI's GPT-4 Vision capability for the understanding of course materials, which can be used to process the multimodal input information. When a student submits a response, the system identifies and retrieves the most semantically relevant slide content, creating a direct connection between course materials and the feedback.

\subsubsection{Multimodal Feedback Generation: Text Feedback with Lecture Slides}
SlideItRight utilizes Retrieval-Augmented Generation (RAG) technology to ensure accuracy and relevance of feedback. RAG enhances knowledge-intensive language processing tasks by incorporating external knowledge databases, producing more specific and accurate responses \cite{gao_retrieval-augmented_2024,lewis_retrieval-augmented_2021}. Supported with lecture slides, the system further extends its capability by supplementing knowledge from other modalities that text alone may not capture \cite{chen-etal-2022-murag}. When a student submits a response,the system uses a two-stage matching process to retrieve the three most semantically relevant slides from course materials.
First, the system leverages multimodal understanding capabilities to analyze the textual, visual, and layout features of each course slide, converting them into comprehensive vector representations that capture both semantic and visual information. Second, it matches the question vector against these slide vectors based on semantic similarity to identify the most relevant course materials. The retrieved slide content is then integrated into the feedback generation process, serving as a knowledge base that grounds the AI responses in verified instructional materials.
This design addresses cognitive load management through: (1) enabling parallel processing of information via separate channels (visual instructional material and textual guidance), aligning with multimedia learning principles \cite{MAYER200285}, and (2) presenting only the most relevant slide page rather than entire documents, reducing extraneous cognitive load while preserving essential content~\cite{cognitiveload}.

\subsubsection{Personalized Feedback}
The system achieves personalization through sophisticated language generation powered by the GPT-4o model.
By analyzing students' free responses to each question, the system generates individually tailored feedback using a learner-centered prompting strategy grounded in learning science theories. The prompts can be found in our public GitHub repository~\footnote{\label{fn1}\github}.

\subsubsection{Efficient Responding Time}
To enhance learning experience, SlideItRight implements an efficient caching system. The vision understanding of slide content is pregenerated and stored in the database, along with relevant slide page references based on question content and retrieval range. This preprocessing significantly reduces real-time computation needs during feedback generation. Additionally, the system preserves student responses for each question, ensuring that their most recent answers are retained even after page refreshes, eliminating the need for re-input and creating a more seamless learning experience.

The system architecture balances sophisticated AI capabilities with practical usability considerations, focusing on supporting student learning while maintaining high standards of feedback quality and relevance.

\vspace{-2mm}
\section{Experimental Design} 
\vspace{-1mm}

\begin{figure}[h!]
    \centering
    \includegraphics[width=0.85\linewidth]{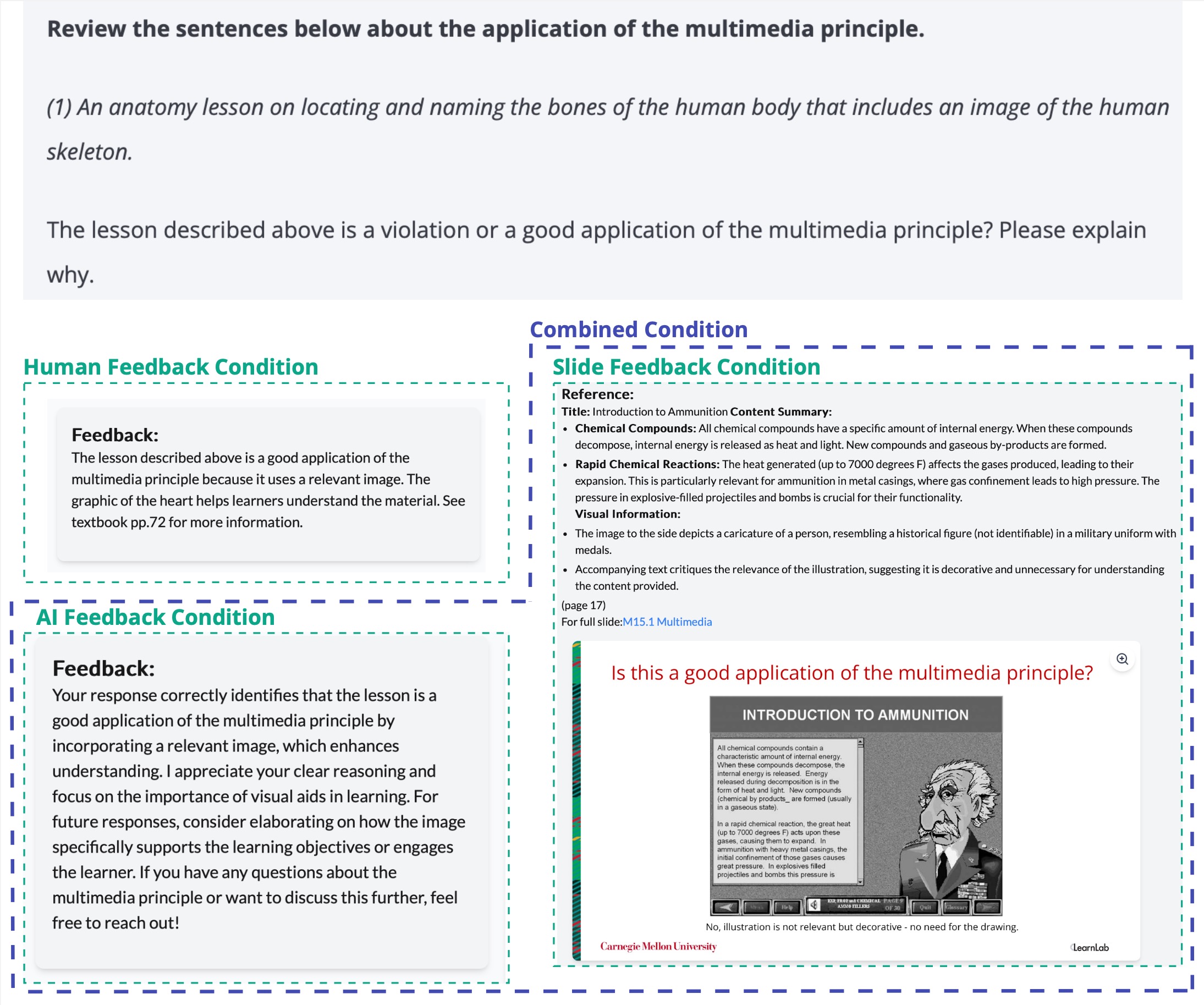}
    \caption{The Feedback Format for Four Experimental Conditions}
    \label{fig:different-feedback-format}
    \vspace{-5mm}
\end{figure}

This study employs a 2×2 experimental design (i.e., human feedback vs. AI feedback and with vs. without relevant slide) to evaluate the impact of LLM-facilitated multimodal feedback on student learning outcomes and perceptions. The study investigates two key variables: (1) the presence of text-based AI-generated feedback and (2) the inclusion of retrieved instructional slide content as supplementary feedback. The research and recruitment procedures were approved by the IRB (Institutional Review Board) at Carnegie Mellon University. 

\subsubsection{Learning Materials}
The instructional content used in this study was derived from a course on instructional design principles in learning engineering and e-learning design. The selected materials focused on the multimedia principle, which is a fundamental concept of e-learning design. The details of the learning materials can be accessed via GitHub repository \textsuperscript{\ref{fn1}}. 

\subsubsection{Conditions} 
The study implemented four different feedback conditions while maintaining consistent instructional content, quiz format, and assessment materials across all conditions. The comparison of format and content differences for each condition is shown in Fig. \ref{fig:different-feedback-format}. Only the feedback modality varied across conditions: Human Feedback Condition provided human-written feedback, Slide Feedback Condition retrieved relevant slide content, AI Feedback Condition delivered AI-generated textual feedback, and Combined Condition showed the combination of AI-generated textual feedback with retrieved slide content.

\subsubsection{Study Procedure}
The study procedure consisted of five sequential steps:

\vspace{-2mm}

\begin{enumerate}
    \item \textbf{Pre-Test Assessment:} Using questions selected from previous course practice, participants completed a baseline knowledge assessment, which included 15 multiple choice questions (MCQs), and an attention check question.
    \item \textbf{Learning Phase I:} Participants engaged with an online learning module on OLI Torus, which incorporated text-based content and instructional slides. The instructional content used in this study was derived from a course on instructional design principles in learning engineering and e-learning design. The selected materials focused on the multimedia principle, which is a fundamental concept of e-learning design. Participants could either begin by viewing the slides or watching a linked video, or alternatively, start answering questions and return to the learning materials as needed to better understand the concepts.
    \item \textbf{Learning Phase II \label{LP2}:} Following the initial learning materials, the participants completed 11 questions (5 MCQs and 5 open-ended questions) for two learning objectives. 
    The open-ended questions were adapted from previously used MCQs, which had been implemented in normal educational practices, to test SlideItRight's personalization capabilities. Participants received immediate feedback after each response based on their assigned condition (Fig. \ref{fig:different-feedback-format}), with MCQ feedback directly embedded in OLI Torus and open-ended question feedback presented through an integrated iFrame component.
    \item \textbf{Post-Test Assessment} To measure learning gains, participants completed a post-test using the same set of questions as the pre-test. Only the attention check question was modified, ensuring comparable difficulty and structure.
    \item \textbf{Post-Survey Evaluation} Participants evaluated their learning experience through a combination of 5-point Likert scale items and open-ended responses. They rated the clarity, engagement, perceived effectiveness, and ease of use of the feedback they received. An attention check question was included to ensure data quality.
\end{enumerate}

\vspace{-3mm}
\subsubsection{Participant Recruitment}
We recruited participants online via Prolific, which is a crowdsourcing platform. The eligibility criteria required participants to be (1) at least 18 years old, (2) English-proficient, and (3) located in the United States. Participants were randomly assigned to four experimental conditions and informed of their right to withdraw at any time. Initially, 100 participants completed the tasks. After filtering for attention check compliance, our final sample consisted of 91 participants (37 women, 54 men) with an average age of 39.5 years. The final distribution included 22 participants receiving human-written textual feedback, 23 receiving AI-written textual feedback, 24 receiving retrieved relevant slide page feedback, and 22 receiving combined AI feedback with relevant slide page feedback. Each participant received \$9 as compensation.

\vspace{-3mm}
\section{Results}
\vspace{-2mm}
\subsection{Learning Gain Measurements: Pre- \& Post- Test Study}

\begin{table}[t!]
\vspace{-6mm}
\caption{Two-Tailed Paired T-Test Results for Learning Gain}\label{tab:ttest}
\centering
    \scriptsize
    \renewcommand{\arraystretch}{1.3}
\begin{tabular}{p{5cm}>{\centering\arraybackslash}m{1cm}>{\centering\arraybackslash}m{1.5cm}}
\hline
\textbf{Feedback Type} & \textbf{t-stat} & \textbf{p-value} \\
\hline
Human Feedback & -2.83 & 0.01010 \\
Relevant Slide Page & -5.00 & 0.00005 \\
AI Feedback & -4.58 & 0.00013 \\
Combined (Slide + AI Feedback) & -3.99 & 0.00067 \\
\hline
\end{tabular}
\vspace{-3mm}
\end{table}

\vspace{-1mm}
To evaluate the effectiveness of SlideItRight in supporting student learning, we analyzed learning gains across different feedback conditions through both pre- and post-test assessments. Learning gains were calculated as the difference between post-test and pre-test scores within the groups, normalized by the maximum possible score. Our analysis revealed that all feedback modalities successfully supported student learning, with each condition showing significant improvements from pre-test to post-test ($p < 0.05$), as shown in Table \ref{tab:ttest}. Students who received human-written feedback demonstrated significant learning gains, as did those who received relevant content from the slide page, AI-generated feedback, and combined slide and AI feedback.

\begin{table}[t!]
\caption{Two-Way ANOVA Results}\label{tab:anova}
\scriptsize
\centering
\renewcommand{\arraystretch}{1.3}
\begin{tabular}{p{3.8cm} >{\centering\arraybackslash}m{1.8cm} >{\centering\arraybackslash}m{1cm} >{\centering\arraybackslash}m{1cm} >{\centering\arraybackslash}m{1.5cm}}
\hline
\textbf{Source} & \textbf{Sum Sq} & \textbf{df} & \textbf{F} & \textbf{p-value} \\
\hline
$C(Feedback Type)$ & 0.0234 & 1.0 & 1.0900 & 0.298 \\
$C(Slide)$ & 0.0114 & 1.0 & 0.5350 & 0.466 \\
$C(Feedback Type):C(Slide)$ & 0.0015 & 1.0 & 0.0687 & 0.794 \\
$Residual$ & 1.95 & 91.0 & - & - \\
\hline
\end{tabular}
\vspace{-8mm}
\end{table}

To understand potential differences between feedback conditions, we then conducted a two-way ANOVA examining the effects of feedback type and slide presence. The analysis revealed no significant main effect for feedback type ($F = 1.09$, $p = 0.298$) or slide presence ($F = 0.53$, $p = 0.466$), as shown in Table \ref{tab:anova}. Furthermore, we found no significant interaction effect between feedback type and slide presence ($F = 0.07$, $p = 0.794$), suggesting that the combination of slides and AI feedback did not produce significant synergistic effects on learning outcomes.

\begin{figure}[b!]
    \vspace{-5mm}
    \centering
    \begin{subfigure}{0.48\textwidth}
        \vtop{ 
        \centering
        \includegraphics[width=\textwidth]{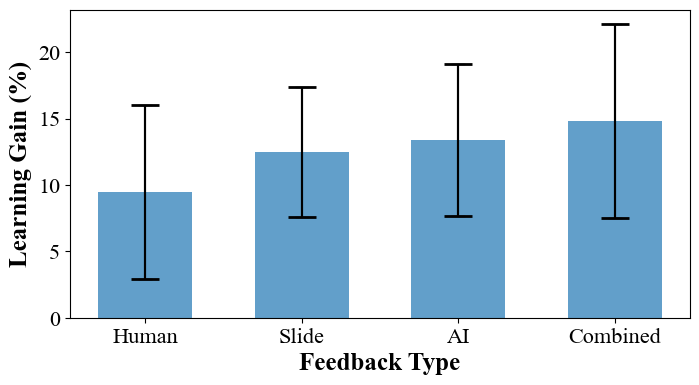}
        \caption{Average Learning Gain with 95\% Confidence Interval}
        \label{fig3a}
        }
    \end{subfigure}
    \hfill
    \begin{subfigure}{0.48\textwidth}
        \vtop{ 
        \centering
        \includegraphics[width=\textwidth]{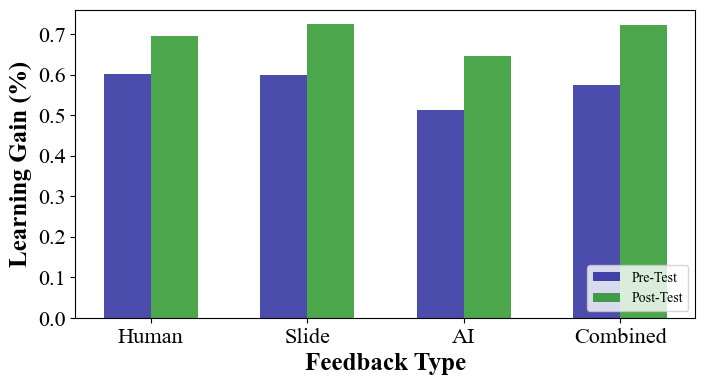}
        \caption{Pre- v.s. Post-Test Scores}
        \label{fig3b}
        }
    \end{subfigure}
    \caption{Comparison of Learning Gains and Test Scores Across Groups}
    \label{fig3}
    \vspace{-5mm}
\end{figure}

While statistical comparisons showed no significant differences between conditions, the data on average learning gains revealed an interesting trend. As shown in Fig. \ref{fig3a}, the combined approach of slides and AI feedback (14.8\%) showed the highest average gain, followed by AI only feedback (13.4\%) and slide-only feedback (12.5\%), with human feedback (9.49\%) showing the lowest average gain. In addition, we performed a one-way ANOVA on pre-test scores ($F = 1.73$, $p = 0.167$), which did not detect significant differences in prior knowledge between the groups. This statistical verification ensures that the observed trends in learning gains were not influenced by the initial differences between the groups. However, given the lack of statistical significance in the comparison of learning gains, these trends should be interpreted cautiously and may warrant further investigation with larger sample sizes.
\vspace{-2mm}

\begin{table}[b!]
\vspace{-6mm}
\caption{Proportion of `Agree' or Higher in the Post-Survey Likert Responses}\label{tab:likert_results}
\centering
\scriptsize
\renewcommand{\arraystretch}{1.25} 
\setlength{\tabcolsep}{5pt}
\begin{tabular}{p{5.1cm}|p{1.3cm}|p{1.3cm}|p{1.3cm}|p{1.3cm}}
\hline
\textbf{Question} & \textbf{Human Feedback (\%Agree)} & \textbf{Slide (\%Agree)} & \textbf{AI Feedback (\%Agree)} & \textbf{Combined (\%Agree)} \\
\hline
\textit{Q1. I am satisfied with my overall learning experience.} \label{q1} & 77.27 & 86.96 & 75.00 & 77.27 \\
\textit{Q2. I feel I gained sufficient knowledge and learning outcomes.} \label{q2} & 81.82 & 86.96 & 75.00 & 77.27 \\
\textit{Q3. Feedback was easy to understand.} \label{q3} & 95.45 & 39.13 & 70.83 & 81.82 \\
\textit{Q4. Feedback for learn-by-doing was helpful.} \label{q4} & 90.91 & 73.91 & 75.00 & 81.82 \\
\textit{Q5. Feedback provided actionable suggestions.} \label{q5} & 68.18 & 52.17 & 87.50 & 59.09\\
\textit{Q6. Feedback encouraged reflection and critical thinking.} \label{q6} & 81.82 & 60.87 & 75.00 & 68.18 \\
\textit{Q7. It was important to me to know whether the feedback was generated by a human or an AI.} \label{q7} &	36.36	& 30.43 &	25.00	&22.73 \\
\textit{Q8. Trust in feedback.} \label{q8} & 81.82 & 65.22 & 50.00 & 68.18 \\

\textit{Q9. Feedback addressed issues in my responses.} \label{q9} & 81.82 & 56.52 & 79.17 & 86.36 \\
\textit{Q10. Feedback was personalized.} \label{q10} & 72.73 & 56.52 & 79.17 & 68.18 \\
\textit{Q11. Feedback motivated me to engage.} \label{q11} & 90.91 & 60.87 & 62.50 & 63.64 \\


\hline
\end{tabular}

{\raggedright Note: This table presents results for questions relevant to the discussion in this paper. The full proportion of ‘Agree’ or higher responses for four questions is detailed via GitHub repository \textsuperscript{\ref{fn1}}. \par}
\vspace{-5mm}
\end{table}

\subsection{Student Perceptions: Post-Survey}
Our analysis examined how different feedback approaches af fected student perceptions, focusing on key qualities such as clarity, personalization, trust, and engagement. The data were collected quantitatively using a 5-point Likert scale (Cronbach's $\alpha$ > 0.90 for all conditions) and qualitatively through open-response questions. Table \ref{tab:likert_results} presents the percentages of students who selected ``\textit{Agree}'' or ``\textit{Strongly Agree}'' in the post-survey for the feedback that they received for the learn-by-doing activities in \hyperref[LP2]{Learning Phase II}. Each feedback method demonstrated distinct strengths and challenges across these dimensions.

\subsubsection{Overall Satisfaction}
Participants reported high satisfaction with the learning experience of four conditions, particularly when feedback was structured and directly applicable to their learning. Slide feedback received the highest satisfaction score (86.96\%), while AI text feedback condition had the lowest (75.00\%) (\hyperref[q1]{Q1}). Similarly, slide feedback was rated the highest in terms of perceived learning gains (86.96\%), followed closely by human feedback (81.82\%) (\hyperref[q2]{Q2}).

\subsubsection{Clarity and Actionability}
Human-written feedback was perceived as the most comprehensible, with 95.45\% of the participants agreeing that it was easy to understand (\hyperref[q3]{Q3}) and 90.91\% finding it helpful (\hyperref[q4]{Q4}). One participant noted that human feedback \textit{``is easy to follow and does not require much effort to understand how to use it''}. Slide feedback received the lowest clarity rating (39.13\%) (\hyperref[q3]{Q3}), with learners requesting \textit{``simple clarity, NOT verbose responses that felt straight out of a dictionary''}. The slide-based approach, while useful as a reference, lacked direct improvement guidance. Although AI feedback (70.83\%) and combined feedback (81.82\%) performed generally well (\hyperref[q3]{Q3}) and was recognized for pinpointing errors quickly, they sometimes struggled with complexity, especially for beginner-level students. A participant criticized that AI feedback \textit{``fails to explain in simple terms for a new learner, and assumes that the user is advanced already and is prepared to read 500-1000 word responses for a single wrong 4 choice answer''}. Future iterations of SlideItRight should focus on simplifying language use while maintaining knowledge depth.

Beyond clarity, effective feedback should provide actionable insights that guide learners toward improvement. AI feedback was rated as the most actionable (87.50\%), followed by human feedback (68.18\%), and slide-based feedback received the lowest score (52.17\%) (\hyperref[q5]{Q5}). The participants commented that the AI feedback \textit{``told me what I was correct about and also what I precisely got wrong helped a lot''} and \textit{``helped me narrow down where my issues are, and what I need to focus on learning''}. Meanwhile, it was pointed out that the learner needed clear explanations on how to revise or improve incorrect answers beyond simply providing references like lecture slides.

\subsubsection{Trust and Reliability}
Human feedback was perceived as the most reliable (81.82\%), followed by slide feedback (65.22\%) (\hyperref[q8]{Q8}). AI feedback struggled with trust (50.00\%) (\hyperref[q8]{Q8}), primarily due to concerns about accuracy and confidence in the responses. One participant noted uncertainty: \textit{``I wasn't entirely sure if the feedback was being nice to me or if my answers were actually good and to some extent that created a sort of crisis of confidence''}. The combined approach received higher trust than AI or slide feedback and showed comparable high relevance level to human feedback. However, consistency remained an issue, as one participant noted: \textit{``Having AI feedback and retrieved slides was helpful, but sometimes they contradicted each other''}. While participants valued AI feedback for its immediacy, overgeneralization and lack of precision reduced trust: \textit{``All the system did was acknowledge my answer and say 'but there's more you could do'... it's simply 'take answer, re-phrase answer back and add 'but you could say more'''}. Addressing this may require integrating more explicit justifications within AI responses or incorporating human oversight in AI-assisted learning.

\subsubsection{Personalized Learning Experience}
AI feedback excelled in personalization rating (79.17\%), outperforming both human (72.73\%) and slide-based feedback (56.52\%) (\hyperref[q10]{Q10}). Participants particularly valued this aspect: \textit{``Personalized feedback is what would drive me to use it quite often as it made me feel more self-sufficient and empowered''}. Another appreciated the combination of correction and encouragement: \textit{``Detailed explanations of incorrect answers along with positive reinforcement pushed me to improve my submissions''}. Additionally, \textit{``the addition of more personalized feedback after each assessment''} is also one of the major suggestions on how to further enhance human feedback.

\subsubsection{Engagement, Motivation, and Cognitive Load}
Human feedback was most effective in maintaining motivation (90.91\%), while participants who received AI feedback (62.50\%) and slide-based feedback (60.87\%) showed notably lower engagement levels (\hyperref[q11]{Q11}). While learners valued AI feedback's detailed explanations, as one participant noted: \textit{``When feedback offers detailed and clear explanations for right and wrong answers, it enhances the learning experience, making it more interactive and beneficial''}, some participants expressed concerns about dependency on such tools, cautioning that \textit{``a tool should only be used when there is a need for it otherwise the tool may become a crutch for the user''}. The combined feedback sometimes overwhelmed students and risked cognitive overload, particularly within the time constraints of the Prolific platform, e.g., a participant stated:\textit{``Too much information at once made it difficult to process everything in a timed setting''}. This time pressure and information density negatively affected participants' motivation to fully engage with the feedback system.

\vspace{-4mm}
\section{Discussion}
\vspace{-2mm}
This study provides valuable insights into the effectiveness and implications of AI-facilitated multimodal feedback in educational settings, and how different feedback modalities impact student learning and engagement.
\subsubsection{Effectiveness of AI-facilitated Multimodal Feedback}

While learning gains were observed in all feedback conditions, the lack of statistically significant differences between modalities suggests that SlideItRight feedback achieved comparable effectiveness to traditional human feedback in supporting student understanding.
Gains trended upward from human to slide to AI, with the combined feedback condition yielding the highest average improvement. However, this trend did not reach significance, possibly due to variability in question difficulty between pre- and post-tests or random noise (e.g., notably lower pretest scores in the AI group). This raises a critical question: if the combined approach integrates the best of both human-like and content-grounded feedback, why didn't it yield superior results? One possibility is that the increased cognitive load, stemming from potentially conflicting or overly dense feedback, diminished the benefit. The human feedback used in this study did not fully adopt a learner-centered structure. Future research should ensure structural alignment across feedback types (e.g., all following learner-centered design principles).
\subsubsection{Student Perceptions and Learning Experience}
Our survey findings revealed distinct advantages and limitations across feedback modalities. Human feedback excelled in clarity and engagement, highlighting the importance of well-structured explanations. While slide-based feedback provided valuable reference materials, it lacked specific guidance for improvement. AI-generated feedback demonstrated strengths in rapid response and personalization, but faced challenges in establishing trust, with students expressing uncertainty about its reliability. The combined approach offered comprehensive learning support, but risked overwhelming students with high information density, particularly when AI feedback and the retrieved slides presented potentially conflicting information, even though the completion time across groups were not significantly different. These insights suggest that, while AI can effectively automate feedback delivery, there is potential for refinement to effectively manage information load through: (1) enhancing explanation clarity \cite{cognitiveload} by providing explanations for specialized terms, (2) building trust mechanisms \cite{NAZARETSKY2025100368} by ensuring AI feedback's explainability with clear rubrics for the questions and establishing appropriate thresholds for when to stop suggesting improvements with phrases like "you could say more"; improving perceived reliability by providing inline references that directly point to specific slide information to build stronger links with reliable course materials, (3) developing adaptive scaffolding strategies \cite{cognitiveload} by implementing multi-level hints, and (4) extending feedback channels \cite{MAYER200285} by incorporating audio options and color coding.
\subsubsection{Implications}
For one thing, our findings support the viability of integrating AI-powered feedback system through embedding as external websites into the page, which could occur in existing online learning platforms (e.g., Moodle, Canvas) that are in use in many universities. For another, educators could benefit from this technology. SlideItRight can serve as a valuable supplement for educators to reduce workload while maintaining feedback quality comparable to human responses, thus freeing time for more complex instructional tasks. Additionally, when instructors struggle to recall specific content locations across extensive course materials, SlideItRight can retrieve specific slides, enhancing the accuracy and specificity of feedback. In light of these, SlideItRight's implementation of Retrieval-Augmented Generation demonstrates the potential for grounding AI feedback in course materials, though opportunities exist for enhancing content retrieval mechanisms to better align with student proficiency levels.

\subsubsection{Limitations}
Several factors limit the generalizability of this study. First, the modest sample size and single-domain context constrain its broader applicability. Second, participants were recruited via an online crowdsourcing platform and compensated, which may have introduced motivation bias, particularly in the depth of engagement with the feedback. Third, the exclusive use of multiple-choice assessments may not fully capture nuanced learning or the system’s impact on student reflection. Future research should more explicitly consider how financial incentives and online environments influence participant behavior, and explore system effectiveness across more diverse, classroom-based populations.

\vspace{-4mm}
\section{Conclusion}
\vspace{-2mm}
Our study provides insights into the comparable support for SlideItRight feedback versus human feedback in terms of learning outcomes. While all feedback conditions demonstrated meaningful improvements in student performance, the lack of significant differences between human and AI-generated feedback suggests their complementary potential. Different feedback versions revealed distinct strengths: human feedback excelled in clarity and trust, while AI-generated feedback offered scalability and personalization. The combination with retrieved slide content showed promise, though moderated by cognitive load considerations. Looking forward, SlideItRight's comparable effectiveness and scalability advantages suggest its value in addressing educational feedback challenges, provided careful attention to cognitive load management and adaptive support for learners with varying prior knowledge.


\begin{credits}
\subsubsection{\ackname} 
This research was supported by the Generative AI + Education Tools R\&D Seed Grant at Carnegie Mellon University. CMU GSA/Provost Conference Funding provided funding support to attend this conference.
\end{credits}


%
%
\bibliographystyle{splncs04}
\bibliography{main}

\end{document}